\documentclass[aps,prd,groupedaddress,showpacs,tighten,11pt]{revtex4}
\usepackage{graphicx}
\usepackage{bm}
\usepackage{amsmath}
\usepackage{latexsym}

\begin{document}

\title{Higher dimensional black holes with a generalized gravitational action}

\author{Jerzy Matyjasek}
\email{jurek@kft.umcs.lublin.pl, matyjase@tytan.umcs.lublin.pl}
\affiliation{Institute of Physics, 
Maria Curie-Sk\l odowska University\\
pl. Marii Curie-Sk\l odowskiej 1, 
20-031 Lublin, Poland}
\author{Ma\l gorzata Telecka}                                         
\affiliation{Institute of Physics, 
Maria Curie-Sk\l odowska University\\
pl. Marii Curie-Sk\l odowskiej 1, 
20-031 Lublin, Poland}
\author{Dariusz Tryniecki}
\affiliation{Institute of Theoretical Physics, 
Wroc\l aw University\\
pl. M. Borna 9, 
50-204 Wroc\l aw, Poland}

\date{\today}

\begin{abstract}
We consider the most general higher order corrections to the pure
gravity action in $D$ dimensions constructed from the basis of the
curvature monomial invariants of order 4 and 6, and degree 2 and 3,
respectively. Perturbatively solving the resulting sixth-order equations
we analyze the influence of the corrections upon  a static and
spherically symmetric back hole. Treating the total mass of the system
as the boundary condition we calculate location of the event horizon,
modifications to its temperature and the entropy.  
The entropy is calculated by integrating the local geometric term 
constructed from the derivative of the Lagrangian with respect to the Riemann
tensor over a spacelike section of the event horizon. It is demonstrated
that identical result can be obtained by integration of the first
law of the black hole thermodynamics with a suitable choice of the 
integration constant. We show that reducing coefficients to the Lovelock 
combination, the approximate expression describing entropy becomes exact. 
Finally, we briefly discuss the problem of field redefinition and analyze 
consequences of a different choice of the boundary conditions in which 
the integration constant is related to the exact location of the event 
horizon and thus to the horizon defined mass.

\end{abstract}

%\preprint{}

%%%%%%%%%%%%%%%%%%%%%%%%%%%%%%%%%%%%%%%%%%%%%%%
%%%%%%%%%%%%%%%%%%%%%%%%%%%%%%%%%%%%%%%%%%%%%%%

%%%%%%%%%%%%%%%%%%%%%%%%%%%%%%%%%%%%%%%%%%%%%%%
%%%%%%%%%%%%%%%%%%%%%%%%%%%%%%%%%%%%%%%%%%%%%%%

%%%%%%%%%%%%%%%%%%%%%%%%%%%%%%%%%%%%%%%%%%%%%%%

%%%%%%%%%%%%%%%%%%%%%%%%%%%%%%%%%%%%%%%%%%%%%%%
%%%%%%%%%%%%%%%%%%%%%%%%%%%%%%%%%%%%%%%%%%%%%%%
%%%%%%%%%%%%%%%%%%%%%%%%%%%%%%%%%%%%%%%%%%%%%%%

% insert suggested PACS numbers in braces on next line
% insert suggested keywords - APS authors don't need to do this

%%%%%%%%%%%%%%%%%%%%%%%%%%%%%%%%%%%%%%%%%%%%%%
\maketitle

\section{\label{intro}Introduction}

In recent years gravitation theories with higher derivative terms have
attracted a great deal of attention. Indeed, according to our present
understanding the general relativity is to be treated as the lowest
order term of the effective theory consisting of a series of loop or
string corrections. Typically such corrections are constructed from
higher powers of curvature and their derivatives, and, hence, the
gravitational action $I$ can be written as
\begin{equation}
I=\sum_{k=0}^{m}\alpha_{k}I_{k},  \label{act}
\end{equation}
where $I_{k}$ for $k\geq 1$ is the contribution of operators of
dimension $ 2k ,$  $I_{0}$ is related to the cosmological constant and
$\alpha_{k}$ are arbitrary constants.
Among the higher curvature theories a great deal of activity has been
focused on the Lovelock gravity \cite{Lovelock}. 
In this theory, the Lagrangian $L_{m}$ is constructed 
from the dimensionally extended Euler densities of a $2k$-dimensional manifold
\begin{equation}
{\mathcal{L}_{k}}\,=\,2^{-k}\delta_{a_{1}b_{1}...a_{k}b_{k}}^{c_{1}d_{1}...
c_{k}d_{k}}R_{c_{1}d_{1}}^{\ \ \ \ a_{1}b_{1}}...R_{c_{k}d_{k}}^{\ \ \ \ a_{k}b_{k}},
                    \label{Leon}
\end{equation}
where the generalized $\delta $ function is totally antisymmetric in
both sets of indices.
A $m-$th order Lovelock action, $I_{m},$ is the sum of $m+1$ terms given 
by Eq.~(\ref{Leon}) of ascending complexity
\begin{equation}
I_{m} = \int d^{D}x (-g)^{1/2}L_{m} =\int d^{D}x (-g)^{1/2}\sum_{k=0}^{m}\alpha_{k}{\mathcal{L}_{k}}, 
\label{Llov}
\end{equation}
where $\alpha_{k}$ are arbitrary parameters. If the Lovelock action
includes the terms up to ${\mathcal{L}_{m}},$ the dimension of the spacetime
should satisfy $D\geq 2m+1.$ 

Varying the action functional $I_{m}$ with
respect to the metric tensor one obtains the equations of motion of
pure gravity, which can be schematically written as
\begin{equation}
L^{ab} = \frac{1}{(-g)^{1/2}}\frac{\delta}{\delta g_{ab}} I_{m} =0.
\end{equation}
Originally, $L_{ab}$ has been introduced by Lovelock to demonstrate 
the most general symmetric and divergence-free second rank tensor, 
which can be constructed from the metric and its first and second 
derivatives. Since the
resulting equations of the Lovelock gravity involve at most second
derivatives of the metric it avoids some of the typical problems of the
higher curvature theories \cite{Zwiebach,Zumino}.
Specifically, at the classical level, it avoids singular
perturbations~\cite{Si,Parker,Bhabha} which do not approach their
Einsteinian counterparts as the perturbative expansion parameter is
set to zero, and, when linearized, the Lovelock equations are free of
ghosts. Moreover, the higher-order terms appear quite
naturally as the low- energy limit of the string theory \cite
{Zwiebach,Zumino}. 

At each order ${\mathcal{L}_{k}}$ is a linear combination of the basis curvature
invariants with the particular set of coefficients calculated from Eq.~(\ref{Leon}). 
For example, in the first two (simplest) cases, one has a cosmological constant and
the curvature scalar, for $k=0$ and $k=1$ respectively. At $k=2$ there
are three invariants which are combined into the Gauss-Bonnet term
whereas at $k=3$ the basis has eight members. 

On the other hand, one
may consider the more general curvature terms, with arbitrary coefficients
rather than those inspired by the particular theory. (See for example
\cite{Deser2005,wybourne} and references cited therein). In this case,
the relation between dimension of the spacetime and the order of
higher curvature terms retained in the action functional is lost and
the dynamical equations inevitably involve higher
derivatives of the metric. There is nothing wrong in using such
equations, provided only the physical solutions are selected. However, 
the expected complexity of the resulting equations may be a serious
obstacle in this regard.

The literature devoted to  various aspects of the higher derivative 
gravity is vast indeed. As the examples of such theories one may 
consider the quadratic or higher order gravity 
(see \cite{Stelle2,Stelle1,Steve,Lousto1,Lousto2,Lousto3,Myers:1998gt,Tryn}
and the references cited therein), and, when a negative cosmological
constant is present, the Einstein-Hilbert action in $D=5,$ supplemented
with ${{\rm Riem}}^{2} = R_{abcd}R^{abcd},$ which corresponds to an effective AdS$_{5}$
bulk action~\cite {Nojiri:2001aj}.  

The foregoing discussion indicates that the analyses can be carried 
out in two directions.
First, one can construct and investigate the possible candidate terms
that may appear in $I_{k},$ whereas the second direction of
calculations is to apply the thus constructed equations in the
particular context of black hole physics or cosmology with or without
additional matter fields. The latter approach has been successfully applied in
various contexts in Refs.~\cite
{My,Callan,Lu,Dobado,Nojiri:2001aj,Nojiri:2001ds,Nojiri:2002qn,Cho1,Neup1,Dehghani} 
and in the references cited therein. 

A few words of comment are in
order here. First, it should be observed that we have no information
on $m,$ i. e., the number of terms that should be retained in
Eq.(\ref{act}). Second, and that is closely related to the above
observation, it is really desirable and perhaps unavoidable to
construct the solutions of the dynamical equations which reflect the
nature of their derivation. Finally, it should be observed that
because of complexity of the problem the full system of equations is
probably intractable analytically and one has to construct either
approximate solutions or refer to the numerical methods.

In this note we shall explore the second possibility and
perturbatively solve the equations resulting from the variation of the
$D-$dimensional action (\ref{act}) without a cosmological
constant about the Tangherlini black hole \cite{tangh,Myers1986}. In
doing so the class of solutions are restricted to the admissible ones.
We shall assume that the total action functional $I$ is the sum of the
first three nonvanishing terms, $I_{k}$, constructed from the basis
curvature invariants. That is, we assume arbitrary coefficients rather
than those inspired by the particular theory. The results of this
paper generalize results of Lu and Wise \cite{Lu} \ and may be though
of as a partial generalization of the analogous results obtained
within the framework of the Lovelock gravity \cite
{Bou,Wheeler1,Wheeler2,Cai,Si}.

\section{\label{eqs}Equations}

We shall consider the action functional being a sum of the terms
(conventions are $R_{ab}=R_{\,\,\,acb}^{c}\sim \partial _{c}\Gamma _{ab}^{c}$
, signature $-,+,+,+$)

\begin{equation}
I_{1}=a\int d^{D}x(-g)^{1/2}R,  \label{i1}
\end{equation}
\begin{equation}
I_{2}=\int d^{D}x(-g)^{1/2}\left(
b_{1}R^{2}+b_{2}R_{ab}R^{ab}+b_{3}R_{abcd}R^{abcd}\right)  \label{i2}
\end{equation}
and 
\begin{align}
I_{3}& =\int d^{D}x(-g)^{1/2}\left(
c_{1}R^{3}+c_{2}RR_{a}^{b}R_{b}^{a}+c_{3}RR_{ab}^{\ \ cd}R_{cd}^{\ \
ab}+c_{4}R_{a}^{b}R_{c}^{d}R_{bd}^{\ \ ac}\right.  \notag \\
& +\left. c_{5}R_{a}^{b}R_{b}^{c}R_{c}^{a}+c_{6}R_{a}^{b}R_{bc}^{\ \
de}R_{de}^{\ \ ac}+c_{7}R_{ab}^{\ \ cd}R_{cd}^{\ \ ef}R_{ef}^{\ \
ab}+c_{8}R_{ce}^{\ \ ab}R_{af}^{\ \ cd}R_{bd}^{\ \ ef}\right) ,  \label{i3a}
\end{align}
where $a=(16\pi G_{D})^{-1}$ and $G_{D}$ is Newton's constant. That is, we
will restrict ourselves to scalar terms of order 2, 4 and 6 belonging to
classes $\mathcal{R}_{2,1}^{0}$, $\mathcal{R}_{4,2}^{0}$ and $\mathcal{R}
_{6,3}^{0}$, respectively \cite{wybourne}. In the course of the calculations
we shall assume that the coefficients $b_{i}$ and $c_{k}$ satisfy $
\left\vert b_{i}\right\vert /a<<1$ and $\left\vert c_{k}/b_{i}\right\vert
<<1 $ for $i=1,..,3$ and $k=1,...,8$, respectively.  The case $b_{i}\sim
c_{i} $ can be easily obtained from the former one simply by relegating the
terms proportional to $b_{i}b_{j}$ from the resulting expressions.

It should be noted that depending on the dimension $D$ the curvature terms
may be subjected to additional relations \cite{wybourne}. Moreover, for a
static and spherically symmetric line element we have additional vanishing
combination of the elements of the curvature basis with 
the coefficients depending on $D$ 
\cite{Deser2005}.

Although it is possible to adopt (with small modifications) the results presented
in Refs. \cite{j1,j2,olo}, here we proceed differently and use the Weyl method~ 
\cite{Deser2003,Deser2004a,Deser2004}. The line element describing the
static, spherically symmetric $D=d+2$-dimensional geometry may by cast into
the form 
\begin{equation}
ds^{2}=-f^{2}(r)dt^{2}+h^{-2}(r)dr^{2}+\frac{\delta _{ij}+x_{i}x_{j}}{1-x^{2}
}r^{2}dx^{i}dx^{j}\hspace{1cm}i,j=2,...,d+1,  \label{line_el}
\end{equation}
where $x_{i}$ are the coordinates covering maximally symmetric $d$
-dimensional space. The components of Riemann tensor, the basic ingredients of
our calculations, are simply 
\begin{equation}
R_{ij}^{\ \ km}=r^{-2}(1-h^{2})\left( \delta _{i}^{k}\delta _{j}^{m}-\delta
_{i}^{m}\delta _{j}^{k}\right),
\end{equation}
\begin{equation}
R_{ir}^{\ \ jr}=-r^{-1}hh^{\prime }\delta _{i}^{j},
\end{equation}
\begin{equation}
R_{tr}^{\ \ tr}=-f^{-1}h(f^{\prime }h)^{\prime }
\end{equation}
and 
\begin{equation}
R_{ti}^{\ \ tj}=-r^{-1}f_{1}h^{2}f^{\prime }\delta _{i}^{j},
\end{equation}
where the prime denotes differentiation with respect to the coordinate $r.$ Upon
inserting the line element into $I$ and subsequently varying the thus
obtained reduced action with respect to the functions $f$ and $h,$ one
obtains rather complicated system of equations (not displayed here), which
may be further simplified with the substitution 
\begin{equation}
f^{2}(r)=e^{2\psi (r)}\left( 1-\frac{2M(r)}{r^{d-1}}\right)  \label{f_fun}
\end{equation}
and 
\begin{equation}
h^{2}(r)=1-\frac{2M(r)}{r^{d-1}}  \label{h_fun}.
\end{equation}
Except for certain combinations of the numerical coefficients $b_{i}$ and $
c_{k}$ leading to the dimensionally extended Euler densities, the resulting
equations of motion are still too complicated to be solved exactly.
Fortunately, one can easily devise the perturbative approach to the problem,
treating the higher derivative terms as small perturbations. 

Now, in order
to simplify calculations and to keep control of the order of terms in
complicated series expansions, we shall introduce another (dimensionless)
parameter $\varepsilon ,$ substituting $b_{i}\rightarrow \varepsilon b_{i}$
and $c_{i}\rightarrow \varepsilon ^{2}c_{i}.$ We shall put $\varepsilon =1$
in the final stage of calculations. For the unknown functions $M(r)$ and $
\psi (r)$ we assume that they can be expanded as 
\begin{equation}
M(r)=\sum_{i=0}^{m}\varepsilon ^{i}M_{i}(r)+O(\varepsilon ^{m+1})
\end{equation}
and 
\begin{equation}
\psi (r)=\sum_{i=1}^{m}\varepsilon ^{i}\psi _{i}(r)+O(\varepsilon ^{m+1}).
\end{equation}
The system of differential equations $M_{i}(r)$ and $\psi _{i}(r)$ is to be
supplemented with the appropriate, physically motivated boundary conditions.
First, it seems natural to demand 
\begin{equation}
M(r_{+})=\frac{r_{+}^{d-1}}{2},  \label{first_type}
\end{equation}
or, equivalently, $M_{0}\left( r_{+}\right) =r_{+}^{d-1}/2$  and $M_{i}\left(
r_{+}\right) =0$ for $i\geq 1$, where $r_{+}$ denotes the exact location of
the event horizon. Such a choice leads naturally to the horizon defined
mass, 
\begin{equation}
\mathcal{M}_{H}=\frac{d\omega _{d}}{16\pi G_{D}}r_{+}^{d-1}.  \label{hor_def}
\end{equation}
On the other hand, however, one can relate the additive integration constant
with the total mass of the system as seen by a distant observer. Indeed,
analysis carried out in a weak field regime indicates that the constant of
integration $C$ is related to the mass $\mathcal{M}$ according to the
formula 
\begin{equation}
C=\frac{8\pi G_{D}}{d\omega _{d}}\mathcal{M},  \label{adm}
\end{equation}
where 
\begin{equation}
\omega _{d}=\frac{2\pi ^{(d+1)/2}}{\Gamma ((d+1)/2)}
\end{equation}
is the area of a unit d-sphere. For the function $\psi (r)$ we shall always
adopt the natural condition $\psi (\infty )=0$. Since the results obtained
for each set of boundary conditions are not independent and one can easily
transform solution of the first type into the solution of the second type
(and vice versa), we shall concentrate on the boundary conditions of the
second type. A brief discussion of  the consequences of the boundary
conditions of the first type will be given at the end of the paper.

%%%%%%%%%%%%%%%%%%%%%%%%%%%%%%%%%%%%%%%%%%%%
%%%%%%%%%%%%%%%%%%%%%%%%%%%%%%%%%%%%%%%%%%%%

\section{\label{pert}Perturbative solution}

%%%%%%%%%%%%%%%%%%%%%%%%%%%%%%%%%%%%%%%%%%%%
%%%%%%%%%%%%%%%%%%%%%%%%%%%%%%%%%%%%%%%%%%%%
One expects, on general grounds, that the terms proportional to the
coefficients $c_{1},$ $c_{2}$, $c_{4}$ and $c_{5}$ do not contribute to the
solution$.$ Integration of the zeroth-order equations yields, as expected,
the Tangherlini solution with $M_{0}(r)=C,$ whereas the first-order
equations give 
\begin{equation}
M_{1}(r)=-\frac{2C^{2} b_{3}}{a\,r^{d+1}}(d-2)(d-1)  \label{em1}
\end{equation}
and $\psi _{1}=0.$ 
Integration of the second-order
equations, although straightforward, yields much more complicated results: 
\begin{align}
M_{2}(r)& =\frac{C^{3}(d-1)}{ar^{2+2d}}\left[
4c_{3}d(5+8d+3d^{2})-2c_{6}(3-5d^{2}-2d^{3})-2c_{7}(10+16d-21d^{2}-7d^{3})
\right.   \notag \\
& +\frac{1}{2}c_{8}(2+25d-24d^{2}-7d^{3})-\frac{8b_{1}b_{3}}{a}(d-2)(d+1)(3d+5)-
\frac{16b_{2}b_{3}}{a}(d-2)(d+1)(d+2)  \notag \\
& \left. -\frac{8b_{3}^{2}}{a}(d-2)(4d^{2}+19d+9)\right] -\frac{
C^{2}(d^{2}-1)\left( d+1\right) }{a\,r^{d+3}}\left[
8c_{3}d-c_{6}(2-3d)+12c_{7}(d-1)\right.   \notag \\
& -\left. 3c_{8}(d-1)-\frac{16 b_{1}b_{3}}{a}(d-2)-\frac{12b_{2}b_{3}}{a}
(d-2)-\frac{32b_{3}^{2}}{a}(d-2)\right]   \label{em2}
\end{align}
and 
\begin{align}
\psi _{2}(r)& =-\frac{C^{2}(d^{2}-1)}{a\,r^{2d+2}}\left[
4c_{3}d(3+2d)-2c_{6}(1-2d-d^{2})-6c_{7}(2-2d-d^{2})\right.   \notag \\
& \left. +\frac{3}{2 }c_{8}(2-3d-d^{2})-\frac{8b_{1}b_{3}}{a}(d-2)(2d+3)-\frac{
8b_{2}b_{3}}{a}(d-2)(d+2)-\frac{8b_{3}^{2}}{a}(d-2)(2d+5)\right]. 
\label{psi2}
\end{align}
If $b_{i}\sim c_{k}$ then the terms involving $b_{i}$ in $M_{2}(r)$ and $
\psi _{2}(r)$ should be omitted.
Note that for $D=4$ ($d=2$) the function $M_{1}(r)$ is identically zero,
whereas in $M_{2}(r)$ and $\psi_{2}(r)$ all the terms proportional to
$b_{i}b_{j}$ ($i,j=1,2,3$) are absent.
It is because the Kretschmann scalar, $R_{abcd}R^{abcd},$ can be 
relegated from the action as the functional derivative of the
Gauss-Bonnet term with respect to the metric is zero. 

The line element (\ref{line_el}) with (\ref{em1}-\ref{psi2}) provides the
most general solution to the problem. Since the partial results referring to
a particular dimension and/or the concrete form of the action  exist
in the literature, it is worthwhile to compare them with our solution. For $
D=4$ our solution reduces to that which can be easily constructed
by integrating equations derived by Lu and Wise \cite{Lu}. Similarly, retaining
in the action functional (\ref{i3a})\ only the term proportional to $c_{7}$
and making substitutions $C=M/M_{P}^{2}$, $a=M_{P}^{2}/16\pi $ and $
c_{7}=\alpha /M_{P}^{2}$, where $M_{P}$ is the Planck mass and $\alpha $ is
a coupling constant, one obtains results presented in Ref. \cite{Dobado},
however, with one reservation: We have not observed any difference between
the values of the radii at which $g_{00}$ and $g^{11}$ vanish, that is, of
course, in concord with the general form of the adopted line element.

Before proceeding further let us compare our solutions to the
analogous solutions of the Lovelock gravity. First, let us introduce $
a=\alpha _{1},$ $b_{i}=\alpha _{2}{\tilde{b}}_{i}$ and $c_{i}=\alpha _{3}{
\tilde{c}}_{i},$ where $\alpha _{2}$ and $\alpha_{3}$ are the numerical 
coefficients that stand in front of the Gauss-Bonnet and the third-order 
term of the Lovelock Lagrangian, whereas $\tilde{b}_{i}$ and $\tilde{c}_{k}$
are the coefficients that give rise to the Lovelock terms.
Numerically one has $\tilde{b}_{1}=\tilde{b}_{3}=-\tilde{b}_{2}/4=1$ and 
$\tilde{c}_{1}=1,
\tilde{c}_{2}=-12,\tilde{c}_{3}=3,\tilde{c}_{4}=24,\tilde{c}_{5}=16,\tilde{c}
_{6}=-24,\tilde{c}_{7}=2,\tilde{c}_{8}=-8.$ Inserting the above coefficients
into the solution, after massive simplifications, one obtains  $
h^{2}\left( r\right) =f^{2}\left( r\right) $ (or, equivalently $\psi (r)=0$)
and 
\begin{align}
h^{2}(r)& =1-\frac{16\pi G_{D}{\cal M}}{d\,\omega _{d}r^{d-1}}+4096\frac{\pi
^{3}G_{D}^{3}{\cal M}^{2}}{d^{2}\omega _{d}^{2}r^{2 d}}\alpha _{2}\left[ 1 - 512\frac{
\pi ^{2}G_{D}^{2}{\cal M}}{d\omega _{d}r^{d+1}}(d-2)(d-1)\right] (d-2)(d-1)  \notag \\
& +\alpha _{3}65536\frac{\pi ^{4}G_{D}^{4}{\cal M}^{3}}{d^{3}\,\omega
_{d}^{3}r^{3d+1}}(d-4)(d-3)(d-2)(d-1).  \label{h_lov}
\end{align}

On the other hand, although the action of the Lovelock gravity looks rather
complicated it is possible to construct an exact solution describing static and
spherically symmetric configuration~\cite{Wheeler2,Bou,Wheeler1}. 
Such a solution can be found,
after the substitution $f^{2}=h^{2}=1-r^{2}F(r),$ by solving for real roots
of the $m-th$ order polynomial equation 
\begin{equation}
\sum_{k=0}^{m}\hat{c}_{k}F^{k}=\frac{2C}{r^{d+1},}  \label{exp_lov}
\end{equation}
where the coefficients $\hat{c}_{i}$ are given by 
\begin{equation}
\hat{c}_{0}=\frac{\alpha _{0}}{a}\frac{1}{d(d+1)},\hspace{5mm}\hat{c}_{1}=1
\end{equation}
and 
\begin{equation}
\hat{c}_{k}=\frac{\alpha _{k}}{a}\prod_{n=3}^{2k}(d+2-n)\,\,\,\mathrm{for}
\,\,k>1.
\end{equation}
Note that if the cosmological constant is taken to be zero 
then $\alpha_{0} =0.$
Finally, assuming $\alpha _{2}=b\sim \varepsilon $ and $\alpha _{3}=c\sim
\varepsilon^{2}$ and expanding Eq.~(\ref{exp_lov}) in powers of $\varepsilon,$ 
one easily reproduces Eq.~(\ref{h_lov}).

%%%%%%%%%%%%%%%%%%%%%%%%%%%%%%%%%%%%%%%%%%%%%%%%%%
%%%%%%%%%%%%%%%%%%%%%%%%%%%%%%%%%%%%%%%%%%%%%%%%%%

\section{\label{temp_sect}Temperature and entropy}

%%%%%%%%%%%%%%%%%%%%%%%%%%%%%%%%%%%%%%%%%%%%%%%%%%%
%%%%%%%%%%%%%%%%%%%%%%%%%%%%%%%%%%%%%%%%%%%%%%%%%%%

The approximate location of the event horizon of the black hole solution
derived in the previous section is given by 
\begin{align}
r_{+}& =(2C)^{1/(d-1)}-\frac{d-2}{a}b_{3}(2C)^{-1/(d-1)}-\frac{1}{a}
(2C)^{-3/(d-1)}\big\{c_{3}(d^{2}-1)d+\frac{1}{2}c_{6}(d^{2}-1)(d-1)  \notag \\
& -\frac{1}{2}c_{7}[2-(d-1)(5d-4)d]+\frac{1}{8}c_{8}(10-13d+12d^{2}-5d^{3})-\frac{2}{a}
(b_{1}b_{3}+b_{2}b_{3})(d^{2}-1)(d-2)  \notag \\
& +\frac{1}{2a}b_{3}^{2}(d-2)(4+6d-13d)\big\}.  \label{rh}
\end{align}
For $D=4,$ the first order terms are absent and one easily reproduces Lu-Wise
result 
\begin{equation}
r_{+}=2GM\left[ 1-\frac{\pi }{G^{3}M^{4}}\left( 6c_{3}+5c_{7}+\frac{3}{2}
c_{6}-c_{8}\right) \right] .
\end{equation}

As is well-known, the Hawking temperature, $T,$ can be easily calculated
from the metric tensor without referring to the field equations. The
standard method of obtaining $T$ relies on the Wick rotation. The
Euclidean line element has no conical singularity provided the time
coordinate is periodic with a period $P$ given by 
\begin{equation}
P=4\pi \lim_{r\rightarrow r_{+}}\left( g_{tt}g_{rr}\right) ^{1/2}\left( 
\frac{d}{dr}g_{tt}\right) ^{-1}.  \label{period}
\end{equation}
Its reciprocal is identified with the black hole temperature, which, in the
case on hand, reads 
\begin{align}
T& =\frac{d-1}{4\pi }(2C)^{-1/(d-1)}-\frac{(d-2)(d-1)d}{4\pi a}
b_{3}(2C)^{-3/(d-1)}  \notag \\
& +\frac{d-1}{4\pi a}(2C)^{-5/(d-1)}\{-\frac{c_{7}}{2}(d-4)[2+(d-2)(d-1)d] 
\notag \\
& +\frac{c_{8}}{8}(d-4)[4+(d-5)(d-1)d]-\frac{b_{3}^{2}}{2a}
(d-2)^{2}(4+7d-4d^{2})\} . \label{temp}
\end{align}
It should be noted that the Hawking temperature does not depend on $b_{1},$ $
b_{2},$ $c_{3}$ and $c_{6}.$ This behavior can be traced back to the
possibility of removing curvature terms proportional to this very
coefficients by means of the appropriate field redefinition. Such a
redefinition certainly modifies equations of motion of the test particles
but should not modify the temperature, which, in turn, is to be measured at
infinity. On the other hand, the horizon defined mass leads to the expression
for temperature which depends on the full set of parameters. The reason is
that the horizon defined mass is not the mass measured at infinity. We shall
return to this problem later.

From Eq.~(\ref{temp}) one sees that the heat capacity
$
C_{BH} =\partial {\cal M}/{\partial T}
$
calculated for the Tangherlini black hole
is given by
\begin{equation}
C_{BH}^{T} = -\frac{d \omega_{d}}{4 G_{D}}\left(\frac{d-1}{4\pi T} \right)^{d}
\end{equation}
and is always negative. That means that for $d>1$ such a black hole is thermodynamically 
unstable, i.e., it increases its temperature when radiating. 
On the other hand, for $ T >> 1 $  the higher derivative corrections can modify this
behaviour, and, depending on the signs and values of the 
coupling constants they can  give a nonnegative
contribution to the total heat capacity as can be easily inferred from 
\begin{equation}
C_{BH} =C_{BH}^{T} + \Delta C_{BH},
\end{equation}
where
\begin{eqnarray}
\Delta C_{BH}& =& \frac{\omega_{d}}{4 a G_{D}}(d-3)(d-2)
d^{2}\left(\frac{d-1}{4\pi T} \right)^{d-2} b_{3} +
 \frac{d \omega_{d}}{8 a G_{D}}
 (d-5)(d-4)\left(\frac{d-1}{4\pi T} \right)^{d-4} \nonumber \\
&&\times\left\{ 
      \frac{1}{a}(d-2)^{2}(d+1)^{2}b_{3}^{2}  + \frac{1}{4}\left[ 8 + 4 (d-a)(d-1) d \right] c_{7} 
- \frac{1}{4}\left[ 4-(d-5)(d-1)d \right] c_{8} 
                                                 \right\}.\nonumber\\
                                     \label{heat}
\end{eqnarray}
It should be noted, however, that the validity of Eq.~(\ref{heat}) 
has its limitations.
Indeed, for small black holes the effect of the back reaction
should be taken into account and the perturbative approach fails to be
accurate. Similarly, since the black hole is treated as 
a system to which a thermal description applies, 
one must require that the change of the Hawking temperature
caused by the emission of a single quantum of radiation is small.
As discussed in Ref~\cite{Wilczek1,Wilczek2} the condition for 
the  thermal description to be self-consistent is  
\begin{equation}
|\frac{\partial T}{\partial {\cal M}}| << 1.
                          \label{Wilczek}
\end{equation}
Now, making use of (\ref{temp}) one has
\begin{eqnarray}
 \frac{\partial T}{\partial {\cal M}} & =& 
\frac{4G_{D}}{d \omega_{d}} (2 C)^{-d/d-1} +
\frac{12 G_{D} b_{3}}{a d \omega_{d}}(2C)^{-(d+2)/(d-1)} d (d-2) \nonumber \\
&-&\frac{10 b_{3}^{2}}{a^{2}d \omega_{d}}(d-2)^{2}(4 d^{2}-7d -4)(2C)^{-(d+4)/(d-1)}
\nonumber \\
&+&\frac{5}{2a d \omega_{d}}(d-4)\left[ 4\left( d^{3} -3d^{2}+2d+2\right)c_{7} - 
\left(d^{3}-6d^{2}+5d-4 \right)d_{8} \right](2C)^{-(d+4)/(d-1)},\nonumber \\
\end{eqnarray}
                           \label{willcz}
and, therefore, one concludes that mini black holes
violate the condition (\ref{Wilczek}). On the other hand,  
for sufficiently massive black hole the first term in the right 
hand side of the above equation is dominant, and, consequently,
the thermal description as well as the perturbative approach
is legitimate. Such black holes are, however, thermodynamically
unstable and their qualitative behaviour is similar to Schwarzschild
black hole. Inclusion of the cosmological constant, 
angular momentum or electric charge
changes the situation dramatically.

The entropy of the black hole may be calculated using various method.
Two techniques, however,  are especially well suited for calculations
of the entropy in the higher derivative theories. One of them 
is the Wald's Noether
charge approach~\cite{Wald1,Iyer} whereas the second one is based 
on the field redefiniton~\cite{Jac1,Jac2}.
Here we shall follow
the approach, in which $S$ is given in terms 
of the surface integral  over the event horizon~\cite{Vis1,Jac1,Iyer}:
\begin{equation}
S = \frac{1}{4G_{D}}A_{H} + 4\pi \int_{H} \frac{\partial{\cal L}}{\partial R_{abcd}}
\tilde{g}_{ac}\tilde{g}_{bd} \sqrt{ ^{(d)}g}d^{d}x,
\end{equation}
where $\tilde{g}_{ab}$ denotes the metric in the subspace orthogonal 
to the event horizon, and, in the case on hand, $\cal L$ is the sum 
of the  Lagrangians given by Eqs. (\ref{i2}) and (\ref{i3a}).  
It should be noted that to calculate the entropy to the required order 
it suffices to retain in the line element the terms which are 
linear in $\varepsilon.$ 

The calculation of $S$ consists of three steps.
First, it is necessary to express the line 
element in terms of $r_{+},$ which can be 
easily achieved by inverting relation (\ref{rh}). Simple
calculations yield
\begin{align}
C& =\frac{1}{2}r_{+}^{d-1}+\frac{b_{3}}{2a}(d-2)(d-1)r_{+}^{d-3}-\frac{1}{
a^{2}}(d-2)(d-1)(d^{2}-1)\left( b_{1}b_{3}+b_{2}b_{3}+3b_{3}^{2}\right)
r_{+}^{d-5}  \notag \\
& +\frac{1}{2a}(d-1)^{2}(d+1)\left[ c_{3}d+\frac{c_{6}}{2}\left( d-1\right) 
\right] r_{+}^{d-5}-\frac{1}{4a}(d-1)c_{7}\left[ 2+d(d-1)(4-5d)\right]
r_{+}^{d-5}  \notag \\
& +\frac{1}{16a}(d-1)c_{8}\left[ 10-d(13-12d+5d^{2})\right] r_{+}^{d-5}.
\label{cc}
\end{align}
Equally well one may calculate $M(r)$ using the boundary conditions
of the first kind (\ref{first_type}) and take the limit as $r \to \infty.$
Further, one has to calculate 
\begin{equation}
J^{abcd}=\frac{\partial{\cal L}}{\partial R_{abcd}}
\end{equation}
remembering that
$J^{abcd}$  shares all symmetries of the Riemann tensor. 
Finally, performing simple integration (which actually reduces to 
the multiplication 
of the result by the factor $\omega_{d} r_{+}^{d}$),
after some algebra, one gets the desired result:
%%%%%%%%%%%%%%%%%%%%
%%%%%%%%%%%%%%%%%%%%
\begin{eqnarray}
S &=& \frac{A_{H}}{4G_{d}} + 4\pi \omega_{d} r_{+}^d \varepsilon 
\left(-2 b_{1} R + 2 b_{2}\delta_{p}^{\ q} \  R_{q}^{\ p} 
+ 4 b_{3} R_{tr}^{\ \ tr} \right)_{|H}
+ 4 \pi \omega_{d} r_{+}^{d}\varepsilon^{2}\left\{
c_{3} {\rm Riem}^{2} \right.\nonumber \\
&+& \left. 
c_{6}\left[ 2(R_{tr}^{\ \ tr})^{2} + R_{p i}^{\ \ q j} R_{q j}^{\ \ p i}\right] 
+ 12 c_{7} (R_{tr}^{\ \ tr})^{2} + 
3 c_{8}\left[ R_{t i}^{\ \ t j} R_{r j}^{\ \ r i} -  
(R_{tr}^{\ \ t r})^{2}\right]
\right\}_{|H} + (...),  \label{ss}
\nonumber\\
\end{eqnarray}
where ellipsis denote the terms which are omitted as they will not contribute to the 
result. All quantities are to be calculated at the event horizon and
summation is assumed over repeated indices: $p, q =0,1$ ($t,\,r$)  and $i,j =2...d+1.$
%%%%%%%%%%%%%%%%%%%%
%%%%%%%%%%%%%%%%%%%%

Now we are ready to calculate the entropy of the black hole described 
by  Eqs. (\ref{f_fun},\ref{h_fun}) with (\ref{em1}-\ref{psi2}).
Substituting  formulas collected in Appendix into Eq.~(\ref{ss})one has 
\begin{align}
S& =\frac{r_{+}^{d}\omega _{d}}{4G_{D}}+\frac{1}{2aG_{D}}
b_{3}(d-1)dr_{+}^{d-2}\omega _{d}-\frac{\omega _{d}r_{+}^{d-4}}{2a^{2}G_{D}}
(d-2)(d-1)d(d+1)\left( b_{1}b_{3}+b_{2}b_{3}+3b_{3}^{2}\right)   \notag \\
& +\frac{d\left( d-1\right) }{4aG_{D}}\omega _{d}r_{+}^{d-4}\left[
d(d+1)c_{3}+\frac{1}{2}(d^{2}-1)c_{6}+3d(d-1)c_{7}-\frac{3}{4}(d-1)^{2}c_{8}
\right]. \label{entr}
\end{align}
As expected, the higher order corrections to the action modify the standard
relation between $S$ and the area of the event horizon which is valid only
for the Einstein gravity. 

The entropy can also be calculated employing the first law of
thermodynamics 
\begin{mathletters}
\begin{equation}
M=TdS+\sum_{i}\mu _{i}dQ_{i},  \label{term1}
\end{equation}
where $\mu _{i}$ are the chemical potentials corresponding to the conserved
charges $Q_{i}.$ Making use of Eq. (\ref{term1}) one has 
\end{mathletters}
\begin{align}
S& =\int T^{-1}d\mathcal{M}+S_{0}  \notag \\
& =\int T^{-1}\left( \frac{\partial \mathcal{M}}{\partial r_{+}}\right)
_{Q_{i}}dr_{+}+S_{0},
\end{align}
where the integration constant $S_{0}$ does not depend on $r_{+},$ but
possibly depends on the coupling constants and the spacetime dimension.
It should be noted that in the present approach it is necessary to retain
in the line element all the terms proportional to $\varepsilon^{2}$ also.
After some algebra one obtains the expression describing the entropy,
which for $S_{0} =0$ coincides with  the one given by Eq.~(\ref{entr}).

The integration constant can be determined from the physical
requirement that the entropy vanishes when the horizon radius 
shrinks to zero~\cite{Cai,ross}. 
For the Lovelock theory it has been shown that this condition
leads to the results which are identical with those obtained within the 
framework of the Euclidean approach.
For $d>4$ this procedure gives $S_{0} = 0.$ 

The entropy as given by Eq.~(\ref{entr}) is expressed in terms 
of the exact location of the 
event horizon, $r_{+},$  and therefore it depends on the full 
set of the coupling constants.
However, according to our previous discussion, one can easily reduce their 
number by a suitable choice of representation.
Indeed, expressing the entropy in terms of the total mass as seen by
a distant observer (or, equivalently, $C$) reduces the number of 
remaining coupling constants to three. To demonstrate this, let us
substitute $r_{+}$ given by ~Eq.(\ref{rh}) into Eq.(\ref{entr}) and retain the
terms up to second order in $\varepsilon.$
After some rearrangement, one obtains 
 \begin{eqnarray}
S &=& \frac{\omega_{d}}{4 G_{D}}(2 C)^{d/(d-1)} +  
 \frac{\varepsilon\omega_{d}}{4aG_{D}}(2C)^{(d-2)/(d-1)}d^{2}b_{3}+
\frac{ \varepsilon^{2}\omega_{d}}{8 a G_{D}} (2C)^{(d-4)/(d-1)}\left[
c_{7}\left(2+2d-3d^{2}+d^{3}\right)\right. \nonumber \\
&&\left. -\frac{1}{4}c_{8}\left(4+5d-6d^{2}+d^{3} \right)-
\frac{1}{a} b_{3}^{2}(d-2)^{2}(1+2d)\right].
                                 \label{entrmas}
\end{eqnarray}
The entropy described by the above equation depends only on $b_{3},$
$c_{7}$ and $c_{8}$ as expected.

Now, let us compare (\ref{entrmas}) with the analogous result 
constructed by Lu and Wise~\cite{Lu}. 
Putting  in~(\ref{entrmas}) $d=2 $ and $G_{D} =G,$
one has
\begin{equation}
S = 4 \pi G  M^2 + 64 \pi^{2}\varepsilon b_{3} + 
\frac{2 \pi^{2}}{G^{2} M^{2}}\left(4 c_{4} + c_{8} \right)\varepsilon^{2}.
                            \label{entr2}
\end{equation} 
Inspection of~(\ref{entr2}) shows that it contains 
the term proportional to $b_{3},$ which is absent in
the Lu-Wise paper.
This can be easily understood as Lu and Wise ignored the Gauss-Bonnet 
term, which, in four dimensions, 
does not affect black hole solution of the field equations. It affects, 
of course, the action itself and consequently the entropy,
leading to appearance of a constant term in $S$ that is independent of ${\cal M}.$

Finally, let us restrict values of the coefficients to its
Lovelock combinations. After some manipulations it could be shown,
that (\ref{entr}) reduces to a  simple expression 
\begin{equation}
S=\frac{r_{+}^{d}\omega _{d}}{4G_{D}\alpha _{1}}\left[ \alpha _{1}+\frac{
2\alpha _{2}}{r_{+}^{2}}d(d-1)+\frac{3\alpha _{3}}{r_{+}^{4}}d(d-3)(d-2)(d-1)
\right],
\end{equation}
which is identical with the \emph{exact} result obtained from a general
formula \cite{Cai,Jac3} 
\begin{equation}
S=\frac{\omega _{d}r_{+}^{d}}{4G_{D}}\sum_{n=1}^{m}\frac{nd}{d+2-2n}\hat{c}
_{n}(r_{+}^{-2})^{n-1}
\end{equation}
for $m=3.$ 

\section{\label{fin}Final remarks}

So far we have considered the boundary conditions of the second type
only. Now, we shall briefly examine appropriate solution constructed
with the aid of the conditions (\ref{first_type}) and $\psi \left(
\infty \right) =0$. Since both solutions are not independent one can
treat the solution of  the one type as the useful check of the
other. Before we proceed further let us analyse some general features
of the function $M(r).$ Nature of the
problem and our previous analysis suggests that the solution has the form
\begin{equation}
M\left( r\right) =\tilde{M}\left( r\right) +C_{1},
\end{equation}
with $\tilde{M}\left( \infty \right) =0$ and $\tilde{M}\left( r_{+}
\right) =r^{d-1}/2-C_{1}$, and, consequently, the total mass of the
system as seen by a distant observer, $\mathcal{M}$, can be obtained
from
\begin{equation}
M\left( \infty \right) =C_{1}=\frac{8\pi G_{D}}{d\omega _{d}}\mathcal{M},
\end{equation}
where $C_{1}$ is expressed, by construction, in terms of the exact
location of the event horizon. One expects, therefore, that $C_{1}=C$,
where $C$ is given by Eq.~(\ref{cc}), and this equality can be treated as a
consistency check. Repeating calculations order by order with
the new boundary conditions, after some algebra it can be shown that
the function $M\left( r\right) $ can be written as
\begin{eqnarray}
M\left( r\right) &=&-\frac{b_{3}}{2a}\left( d-2\right) \left( d-1\right) 
\frac{r_{+}^{2d-2}}{r^{d+1}}-\frac{b_{3}^{2}}{a^{2}}\left( d-2\right)
^{2}\left( d-1\right) ^{2}\frac{r_{+}^{2d-4}}{r^{d+1}}  \notag \\
&&+A_{1}\left( r\right) \frac{r_{+}^{2d-2}}{r^{d+3}}+A_{2}\left( r\right) 
\frac{r_{+}^{3d-3}}{r^{2d+2}}+C_{1},  \label{M_eh}
\end{eqnarray}
 where $C_{1}$ is given by Eq.~(\ref{cc}),
\begin{eqnarray}
A_{1}\left( r\right) &=&\frac{\left( d-1\right)\left(d+1\right)^{2} }{a}
\left[
-2 d c_{3} + \frac{1}{4}\left(
2-3d\right)c_{6} -3\left( d - 1\right) c_{7} +\frac{3}{4}\left(
d-1\right)c_{8} \right]  \notag \\
&&+\frac{\left( d-2\right) \left( d-1\right)\left(d+1 \right)^{2}}{a^{2}}\left(
3b_{2}b_{3}+4b_{1}b_{3} +8b_{3}^{2}\right),  \label{A1}
\end{eqnarray}
and
\begin{eqnarray}
A_{2}\left( r\right)  &=&\frac{\left( d-1\right) }{a}\left[ \frac{c_{3}}{2}
d\left( d+1\right) \left( 3d+5\right) -\frac{c_{6}}{4}\left( d+1\right)
\left( 3-3d-2d^{2}\right) \right.   \notag \\
&&\left. -\frac{c_{7}}{4}\left( 10+16d-21d^{2}-7d^{3}\right) +\frac{c_{8}}{4}
\left( 2+25d-24d^{2}-7d^{3}\right) \right]   \notag \\
&&-\frac{1}{a^{2}}\left( d-2\right) \left( d-1\right) \left[
2b_{2}b_{3}\left( d+1\right) \left( d+2\right) +b_{1}b_{3}\left( d+1\right)
\left( 3d+5\right) \right.   \notag \\
&&\left. +b_{3}^{2}\left( 9+19d+4d^{2}\right) \right].   \label{A2}
\end{eqnarray}
Similar calculations carried out for the function $\psi (r)$ yield
\begin{eqnarray}
\psi \left( r\right) &=&\frac{2}{a^{2}}\left( d-2\right) \left( d-1\right)
\left( d+1\right) \left[ b_{2}b_{3}\left( d+2\right) +b_{1}b_{3}\left(
2d+3\right) +b_{3}^{2}\left( 2d+5\right) \right] \frac{r_{+}^{2d-2}}{r^{2d+2}
}  \notag \\
&&-\frac{\left( d-1\right) \left( d+1\right) }{a}\left[ c_{3}d\left(
3+2d\right) -c_{6}\left( 1-2d-d^{2}\right) -\frac{3}{2}c_{7}\left(
2-2d-d^{2}\right) \right.  \notag \\
&&\left. +\frac{3}{8}c_{8}\left( 2-3d-d^{2}\right) \right] \frac{r_{+}^{2d-2}
}{r^{2d+2}}.  \label{ps2}
\end{eqnarray}

Since, by assumption, the radius of  the event horizon is treated as
the exact quantity now, the thus derived line element may be easily
employed in construction of the entropy. First, observe that the
Hawking temperature calculated with the aid of the Eq.~(\ref{period})
is given in terms of $r_{+}$ and 
depends on all relevant coefficients. On the other hand the relation
(\ref{temp}) is independent of $b_{1},\,b_{2},\,c_{4}$ and $c_{5}.$
This behaviour can be ascribed to the possibility to remove the
dependence of the line element on this very coefficients at the
expense of modifications of the equations of motion of test particles.
Indeed, it could be demonstrated that by means of  the field redefinition
of the form
\begin{equation}
g_{ab}\rightarrow g_{ab}+\varepsilon A_{ab}^{\left(  1\right)  }
+\varepsilon^{2}A_{ab}^{\left(  2\right)  },
                                           \label{redef}
\end{equation}
where
\begin{equation}
A_{ab}^{\left(  1\right)  }=q_{1}^{\left(  1\right)  }Rg_{ab}+q_{2}^{\left(
1\right)  }R_{ab}
\end{equation}
and
\begin{eqnarray}
A_{ab}^{\left(  2\right)  }  & =&q_{1}^{\left(  2\right)  }R^{2}g_{ab}
+q_{2}^{\left(  2\right)  }RR_{ab}+q_{3}^{\left(  2\right)  }g_{ab}
R_{cdef}R^{cdef}+q_{4}^{\left(  2\right)  }g_{ab}R_{cd}R^{cd}\nonumber\\
& +&q_{5}^{\left(  2\right)  }R_{ac}R_{b}^{c}+q_{6}^{\left(  2\right)
}R_{acde}R_{b}^{\,\,\,\,cde},
\end{eqnarray}
one can remove all the terms in the action
except these proportional to the parameters $b_{3},$ $c_{7}$ and $c_{8}.$
The coefficients $q^{(1)}_{i}$ and $q^{(2)}_{k}$ can be 
determined by solving, at each order of the expansion, the appropriate 
systems of equations.
As the result of the field  redefinition (\ref{redef}),
one obtains two additional terms $R\Box R$ and $R_{ab}\Box R^{ab},$ 
which can also be removed from the action functional. It should
be noted, however, that such terms appear naturally 
in the effective action of the quantized massive fields in a large 
mass limit~\cite{Avra1,
Avra2,Avra3}.

The geodesic equation after the field redefinition becomes 
\begin{eqnarray}
\frac{d^{2}x^{i}}{ds^{2}} &+&\Gamma_{jk}^{i}\frac{dx^{j}}{ds}\frac{dx^{k}
}{ds}+\varepsilon g^{im}\left(  A_{mk;j}^{\left(  1\right)  }-\frac{1}
{2}A_{jk;m}^{\left(  1\right)  }\right)  \frac{dx^{j}}{ds}\frac{dx^{k}}
{ds}\nonumber\\
&+&\varepsilon^{2}g^{im}\left(  A_{mk;j}^{\left(  2\right)  }-\frac{1}
{2}A_{jk;m}^{\left(  2\right)  }\right)  \frac{dx^{j}}{ds}\frac{dx^{k}}
{ds}-\varepsilon^{2}A^{\left(  1\right)  im}\left(  A_{mk;j}^{\left(
1\right)  }-\frac{1}{2}A_{jk;m}^{\left(  1\right)  }\right)  \frac{dx^{j}}
{ds}\frac{dx^{k}}{ds}=0,
\end{eqnarray}
where $s$ is the affine parameter of the original metric.
Now, one can repeat the arguments of Ref.~\cite{Lu}.
Both $\mathcal{M}$ and $T$ can be measured at infinity and do not
depend on the particular form of the equations of motion.
Consequently, the temperature mass relation is independent
of the removed terms. However, to
determine the radius of the event horizon one performs local
measurements and the equations of motion of test particles are
important. 

It could be easily seen that, as expected, the entropy is
precisely the same for both choices of boundary conditions and is
described by the formula (\ref{entr}). Since the calculations for both
types of the boundary conditions have been carried out independently,
this equality may be regarded as the additional consistency check.

\appendix*
\section{}

In this Appendix we list some formulas used in Section \ref{temp_sect}.
The expansions of the components of the Riemann tensor 
and some of its contractions up to the first order in $\varepsilon,$
which are necessary in calculation of the entropy, are given by
\begin{equation}
R_{ti}^{\ \ tj}=R_{ri}^{\ \ rj}=-\frac{\left(  d-1\right)}{2 r_{+}^{2}}\delta
_{i}^{j}+\varepsilon b_{3}\frac{\left(  d-2\right)  \left(  d-1\right)
\left(  d+1\right) }{2a r_{+}^{4}}\delta_{i}^{j}+O\left(  \varepsilon
^{2}\right),
\end{equation}
\begin{equation}
R_{ij}^{\ \ kl}=\frac{1}{r_{+}^{2}}\left(  \delta_{i}^{k}\delta_{j}^{l}-\delta_{i}
^{l}\delta_{j}^{k}\right)  +O\left(  \varepsilon^{2}\right),
\end{equation}
\begin{equation}
R_{tr}^{\ \ tr}=\frac{d\left(  d-1\right)}{2 r_{+}^{2}}-\varepsilon b_{3}
\frac{3\left(  d-2\right)  \left(  d-1\right)  d\left(  d+1\right)}{2a r_{+}^{4}}
+O\left(  \varepsilon^{2}\right),
\end{equation}
\begin{equation}
R_{t}^{t}=R_{r}^{r}=-\varepsilon b_{3}\frac{\left(  d-2\right)  \left(
d-1\right)  d\left(  d+1\right) }{2a r_{+}^{4}}+O\left(  \varepsilon
^{2}\right)
\end{equation}
and
\begin{equation}
R=-\varepsilon b_{3}\frac{\left(  d-2\right)  \left(  d-1\right)  d\left(
d+1\right)}{2a r_{+}^{4}}+O\left(  \varepsilon^{2}\right),
\end{equation}
where $i, j, k, l =2,...,d+1.$

%\bibliography{higher_deriv}

\end{document}